\documentclass[conference]{IEEEtran}

\usepackage[utf8]{inputenc}
\usepackage{amsmath,amssymb}
\usepackage{graphicx}
\usepackage{cite}
\usepackage{booktabs}  
\graphicspath{{./Figures/}}
\usepackage{float}
\usepackage{array}
\usepackage{xurl}
\let\labelindent\relax
\usepackage{enumitem}
\usepackage[colorlinks=true, linkcolor=blue, urlcolor=blue, citecolor=blue]{hyperref}

\begin{document}

\title{A Statistical Evaluation of Indoor LoRaWAN Environment-Aware Propagation for 6G: MLR, ANOVA, and Residual Distribution Analysis}

\author{
  Nahshon Mokua Obiri and Kristof Van Laerhoven\\
  \textit{Dep. of Electrical Engineering and Computer Science (Ubicomp Group), University of Siegen, Germany}\\
  \{nahshon.obiri@student.uni-siegen.de, kvl@eti.uni-siegen.de\}
}

\maketitle

\begin{abstract}
Modeling path loss in indoor LoRaWAN technology deployments is inherently challenging due to structural obstructions, occupant density and activities, and fluctuating environmental conditions. This study proposes a two-stage approach to capture and analyze these complexities using an extensive dataset of 1,328,334 field measurements collected over 6 months in a single-floor office at the University of Siegen’s  Hölderlinstraße Campus, Germany. First, we implement a multiple linear regression (MLR) model that includes the traditional propagation metrics (distance, structural walls) and an extension with proposed environmental variables (relative humidity, temperature, carbon dioxide (CO\textsubscript{2}), particulate matter, and barometric pressure). Using analysis of variance (ANOVA), we demonstrate that adding these environmental factors can reduce unexplained variance by 42.32\%. Secondly, we examine residual (shadow fading) distributions by fitting five candidate probability distributions: Normal, Skew-Normal, Cauchy, Student's \(t\), and Gaussian Mixture Models (GMM) with 2 to 5 components. Our results show that a four-component GMM captures the residual heterogeneity of indoor signal propagation most accurately, significantly outperforming single-distribution approaches. Given 6G’s push for ultra‐reliable, context‐aware communications, our analysis shows that environment‐aware modeling can substantially improve LoRaWAN network design in dynamic indoor IoT deployments.
\end{abstract}


\IEEEpeerreviewmaketitle

\section{Introduction}
\label{sec:intro}

Indoor LoRaWAN deployments present a formidable challenge for signal propagation modeling, owing to complex interactions between distance, wall composition, and ambient environmental variables such as temperature, humidity, and occupant density \cite{cattaniExperimentalEvaluationReliability2017}. These factors lead to multipath fading, non-trivial attenuation, and shadowing effects that remain difficult to capture in conventional log-distance path loss models (LDPLMs) \cite{grubelDenseIndoorSensor2022}. Moreover, domain knowledge imposes strict statistical validity requirements for LDPLMs \cite{xuMeasurementCharacterizationModeling2020}. These include confirming the significance of the path loss exponent through an analysis of variance (ANOVA) test. Additionally, the residual error (or shadow fading) must satisfy the following conditions: \textbf{\textit{(i)}} follow a log-normal distribution, \textit{\textbf{(ii)}} be homoscedastic (demonstrate homogeneity of variance), and \textbf{\textit{(iii)}} be uncorrelated with the independent variables (e.g., distance, wall composition, and environmental factors) and with itself (i.e., free from autocorrelation) \cite{gonzalez-palacioLoRaWANPathLoss2023}. However, many studies either overlook these requirements or report residuals that fail normality tests and hence model unreliability \cite{kimExperiencingLoRaNetwork2019}.

We propose a robust statistical approach designed to overcome the limitations prevalent in current indoor LoRaWAN propagation studies. Our objective is to clearly isolate and quantify the combined effects of distance, structural barriers, and dynamic environmental variables, providing deeper insights into their contributions to indoor signal attenuation. Leveraging a six-month empirical campaign comprising \(1,328,334\) observations collected in an operational office environment, our analysis ensures adherence to statistical validity criteria essential for accurate modeling and practical network optimization. A comprehensive description of this dataset can be found in our previous work \cite{mokuaobiriComprehensiveDataDescription2025}. In this work, our analytical framework consists of the following key components:

\begin{enumerate}[label=\textbf{\textit{(\roman*)}}]
    \item A multiple linear regression (MLR) model that captures distance, wall composition, and ambient environmental metrics of relative humidity, temperature, carbon dioxide (CO\textsubscript{2}), particulate matter (PM\textsubscript{2.5}), and barometric pressure,
    \item An ANOVA procedure to rigorously test for variable statistical significance and
    \item A detailed shadow fading (residual distribution) analysis to check compliance with theoretical assumptions on error normality, autocorrelation, and homoscedasticity.
\end{enumerate}

In real-world indoor deployments, incorporating environmental data is crucial for improving path loss predictions \cite{gonzalez-palacioMachineLearningBasedCombinedPath2023a}, which enables power optimization, accurate localization, and robust network design. Although continuously monitoring these parameters may increase power consumption and require adaptive calibration to cope with dynamic occupancy and Heating, Ventilation, and Air Conditioning (HVAC)-induced variations, these measurements provide actionable insights. By deploying low-cost sensors alongside LoRaWAN nodes, practitioners can enable real-time path loss adjustments and design fade margins based on multimodal shadowing, thereby ensuring energy-efficient and resilient operation. Therefore, our study offers a statistically robust and practically viable toolkit for next-generation indoor LoRaWAN path loss modeling, paving the way for more resilient and energy-efficient deployments in the context of 6G and IoT applications.

\section{Literature Review}\label{sec:literature}

\subsection{Indoor LoRaWAN Path Loss Modeling}

Indoor LoRaWAN systems face an inherently complex propagation environment shaped by a broad spectrum of building materials, room geometries, and obstacles that produce significant multipath and attenuation effects \cite{azevedoCriticalReviewPropagation2024}. While traditional log-distance path loss equations provide a starting point, recent investigations highlight substantial deviations in environments featuring thick walls, multi-story layouts, and corridor-like structures \cite{robles-encisoLoRaZigbee5G2023}. Several refined approaches have thus emerged, such as incorporating multi-wall and multi-floor penalties based on empirical measurements. In large, high-rise buildings, for example, calibrating the path loss exponent and specific wall-attenuation constants can significantly improve predictive accuracy, especially when deployed nodes exhibit varying distances and angles of incidence relative to the gateway \cite{aksoyComparativeAnalysisEnd2024}. No single closed-form model captures the diversity of building configurations. Consequently, recent work integrates classical propagation models with site-specific calibration or multi-parameter regression to better model wall materials and spatial layouts \cite{alkhazmiAnalysisRealWorldLoRaWAN2023}.

\subsection{Environmental Factors in Indoor Propagation}

Beyond static architectural features, ambient environmental factors play a significant role in shaping indoor LoRaWAN performance. During extended measurement campaigns, researchers have observed that fluctuating temperature, humidity, and CO\textsubscript{2} levels often correlate with changes in shadowing and multipath variability \cite{muppalaInvestigationIndoorLoRaWAN2021}. In addition, occupant behavior can periodically disrupt established paths or introduce new reflective surfaces, a phenomenon especially pertinent to open-plan offices where human movement is frequent and unpredictable. Longitudinal datasets collected over days or weeks show that peak occupancy times coincide with reduced signal stability, reinforcing the idea that building inhabitants are effectively time-varying obstacles \cite{grubelDenseIndoorSensor2022}, \cite{harindaPerformanceLiveMultiGateway2022}. For particular application domains, such as environmental monitoring or smart HVAC control, these inter-dependencies between occupant density, microclimate conditions, and signal degradation are more than theoretical curiosities. They represent real operational concerns that necessitate environment-aware path loss models. Consequently, a growing body of work now incorporates sensor inputs such as CO\textsubscript{2} or temperature into regression or machine learning (ML) models, demonstrating noticeable improvements in prediction accuracy for complex indoor deployments.

\subsection{Advanced Statistical Approaches and Model Validation}
In parallel with the move towards more comprehensive predictor sets, recent studies emphasize robust statistical validation techniques to ensure that indoor propagation models extend beyond mere curve fitting. One promising direction is hybrid or adaptive modeling, wherein conventional path loss equations are augmented by machine learning components that learn nonlinear corrections for local conditions \cite{aksoyComparativeAnalysisEnd2024}. Long-duration data collection also enables researchers to conduct detailed residual analyses, examining potential multimodality, heavy tails, or time-varying fluctuations in error distributions and verify key assumptions such as normality and homoscedasticity. In one notable example, a dynamic path loss model employing an enhanced Kalman filter achieved more precise distance estimations by iteratively refining received signal strength indicator (RSSI) measurements to account for fast-fading and sudden obstructions \cite{voAdvancePathLoss2024}. These advanced methods illustrate an ongoing shift toward statistically rigorous frameworks, where factors are systematically selected (e.g., via ANOVA), and diagnostic checks are performed on residuals, aiming to produce robust, predictive models for the ever-changing conditions in indoor LoRaWAN deployments.

\section{Methodology}\label{sec:method}

\subsection{Measurement Experimental Setup}

\begin{figure}[hbt!]
\centering
\centerline{\includegraphics[width=.7\columnwidth]{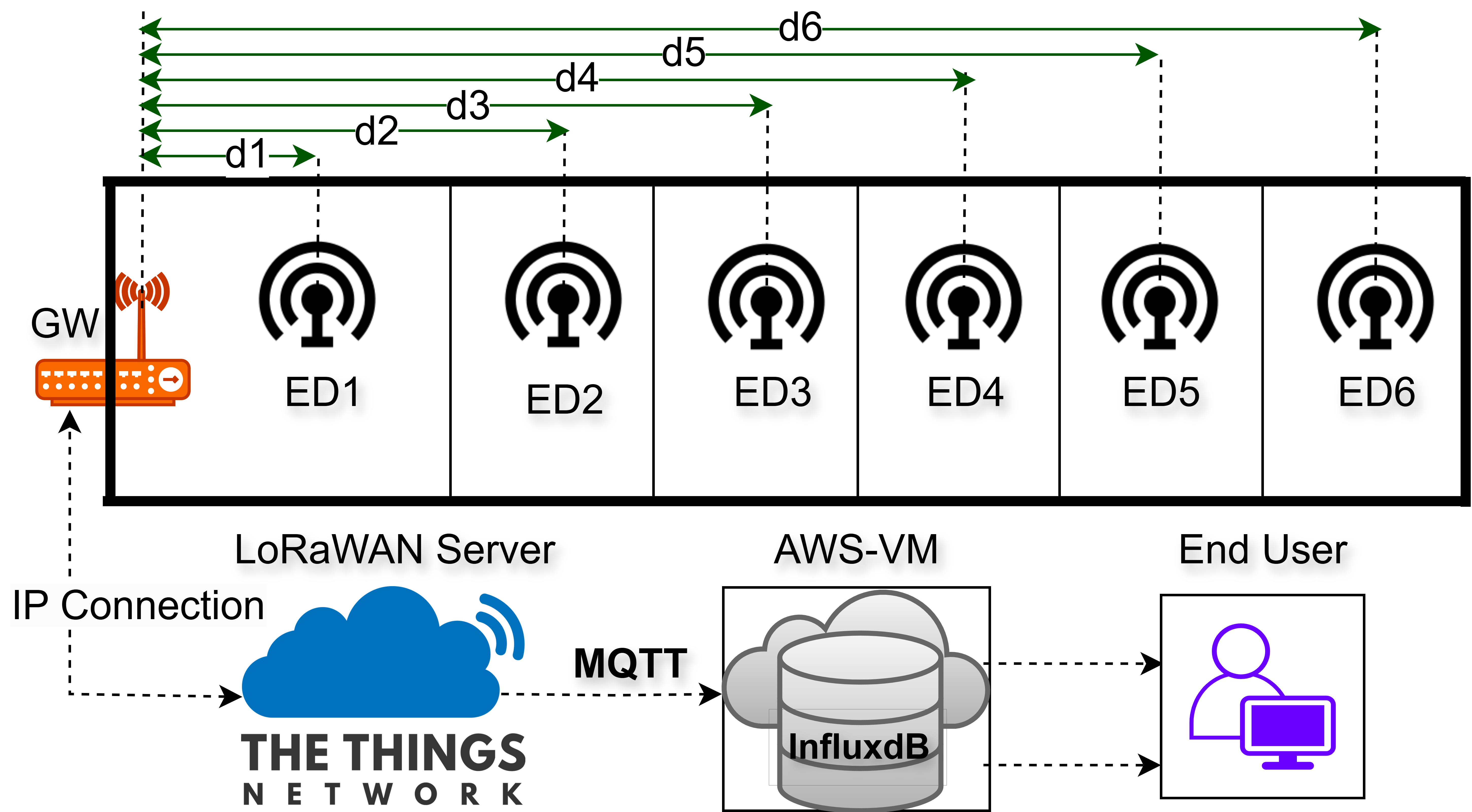}}
\caption{\label{fig: design} {Experimental end devices (EDs) and gateway (GW) deployment layout (not drawn to scale). EDs $ (\text{ED1--ED6}$) and the GW are placed in an indoor environment with brick/concrete and wooden partition walls.}}
\end{figure}

\begin{figure}[hbt!]
\centering
\centerline{\includegraphics[width=.8\columnwidth]{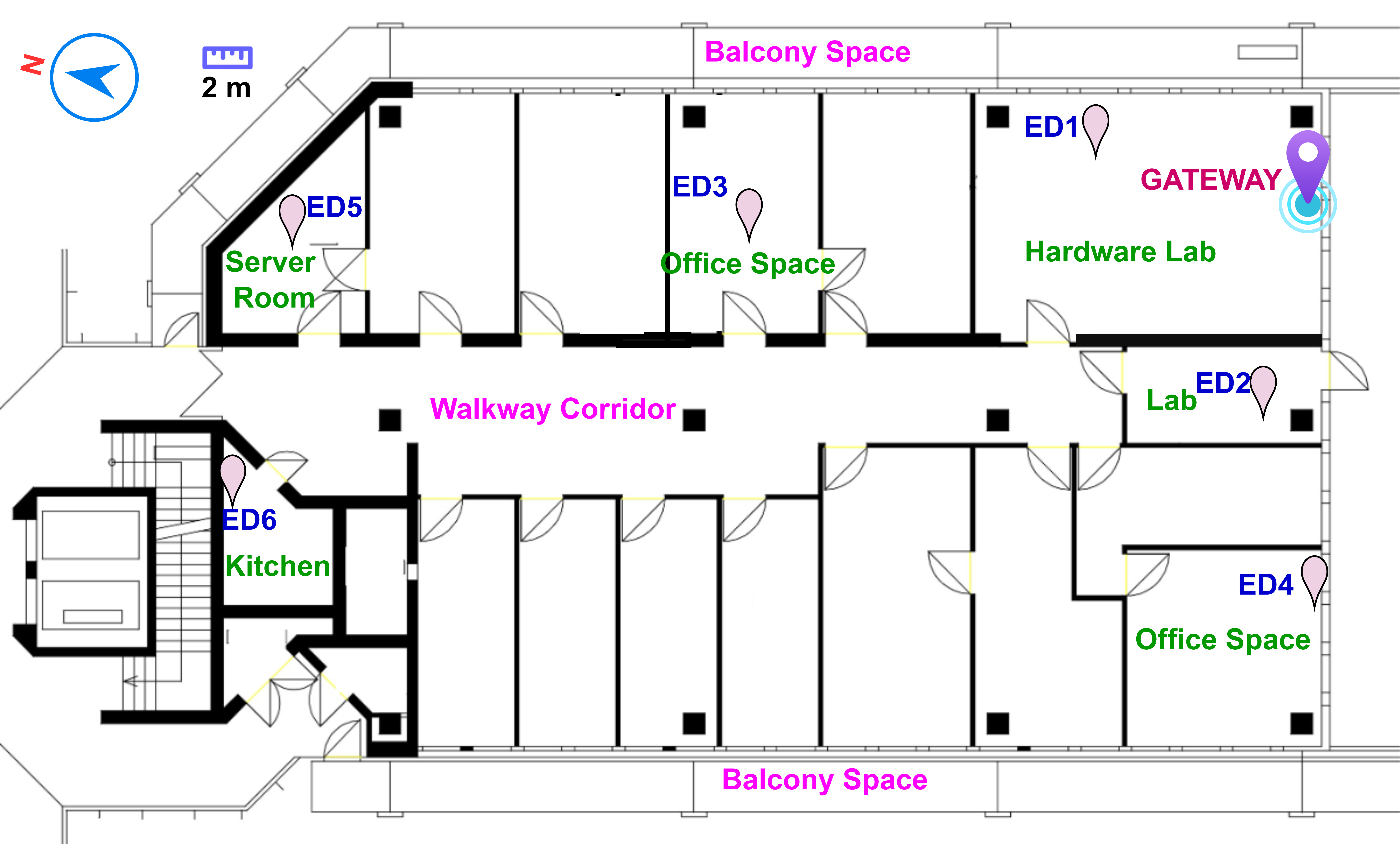}}
\caption{\label{fig: floor_plan} Sensor network deployment layout showing sensor nodes (ED1–ED6) and the gateway (GW) placement in the indoor environment.}
\end{figure}

In this study, we set up a sensor network of \(6\) LoRaWAN end devices (ED1 to ED6), as shown in Figure~\ref{fig: design}. This was deployed on the eighth floor (approximately \(240\,\mathrm{m}^2\), \(290\,\mathrm{m}\) above sea level) of an academic building at the University of Siegen in Germany. The objective was to systematically measure indoor path loss and ambient environmental parameters. We placed the end devices at distances ranging from \(8\,\mathrm{m}\) to \(40\,\mathrm{m}\) from a centrally located Wirnet iFemtoCell indoor gateway, as shown in Figure~\ref{fig: floor_plan}.

This layout aimed to capture a range of realistic indoor propagation conditions, including brick and concrete walls, wooden partitions, and a semi-open corridor. Antenna heights were maintained around $0.8\,\mathrm{m}$ above the floor level for the end devices and $1\,\mathrm{m}$ for the gateway. All signals were forwarded to The Things Network (TTN), where a custom JavaScript decoder parsed the LoRa frames. A Python-based Message Queuing Telemetry Transport (MQTT) subscriber then processed and uploaded the measurements to an InfluxDB database hosted on an Amazon Web Services (AWS) Elastic Compute Cloud (EC2) instance, ensuring continuous time-series logging and automated alerting for data gaps.
 
\begin{figure}[hbt!]
\centering
\centerline{\includegraphics[width=.5\columnwidth]{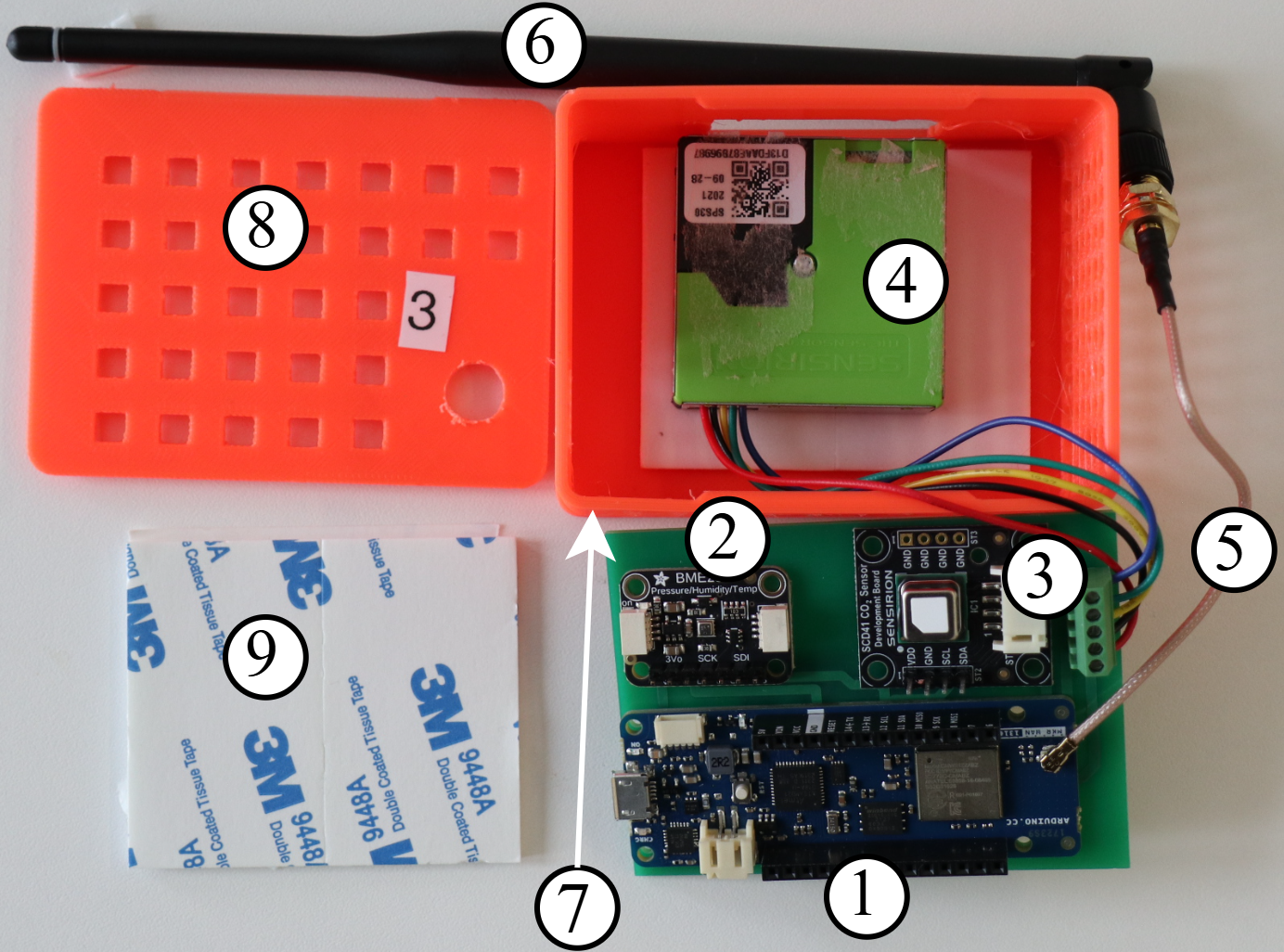}}
\caption{\label{fig: end_device} {End device individual components: (1) Arduino MKR WAN 1310, (2) Adafruit BME280 sensor, (3) Sensirion SCD41 sensor, (4) Sensirion SPS30 sensor, (5) SMA to uFL adapter cable, (6) Rubber duck antenna, (7) 3D-printed casing base, (8) 3D-printed casing lid, (9) Adhesive mounting pads.}}
\end{figure}

Figure~\ref{fig: end_device} shows the assembled end device, which integrates multiple environmental sensors into a custom 3D-printed enclosure. Each end device uses an Arduino MKR WAN 1310 microcontroller with a Murata LoRa radio module configured for the $868\,\mathrm{MHz}$ European band. The hardware is outfitted with a dedicated particulate matter sensor (Sensirion SPS30), a combined temperature-humidity-CO\textsubscript{2} unit (Sensirion SCD41 sensor), and a barometric pressure sensor (Adafruit BME280 sensor). Their data, scaled and packed into an $18\text{-}\mathrm{byte}$ LoRaWAN payload, were transmitted at intervals of $60\,\mathrm{s}$ for $6\,\mathrm{months}$ beginning in late October 2024.
\subsection{Multiple Linear Regression (MLR)}

MLR is a core statistical technique for exploring how multiple predictors \(\{x_1, x_2, \dots, x_n\}\) with \(n>1\) jointly influence a response variable \(y\). Equation \eqref{eq:mlr} represents the expression for MLR:
\begin{equation}
y = \beta_0 + \beta_1 x_1 + \beta_2 x_2 + \dots + \beta_n x_n + \epsilon,
\label{eq:mlr}
\end{equation}
where \(\beta_0, \dots, \beta_n\) denote regression coefficients, and \(\epsilon\) captures residual error \cite{montgomeryIntroductionLinearRegression2012}. Beyond estimating these coefficients, an ANOVA can be performed to confirm the overall statistical significance of the regression and to check whether each predictor contributes meaningfully \cite{farawayPracticalRegressionAnova2002}. To prevent biased parameter estimates and lend greater credibility to both the model’s predictions and subsequent statistical inferences, it is essential to verify that the residuals satisfy the usual MLR assumptions \cite{gonzalez-palacioMachineLearningBasedCombinedPath2023a}, as listed in Section~.\ref{sec:intro}.

\subsection{Model Specification}

This work adopts a log-distance path loss and shadowing model with multiple walls (the COST 231 Multi-Wall Model (MWM) \cite{europeancommissionCOSTAction2311999} ) and additional environmental parameters to capture indoor signal attenuation as influenced by human activities comprehensively. As shown in Eqn.~\eqref{eq:pl-mw-en}, the path loss \(PL\) depends on the distance \(d\), frequency \(f\), the number of walls \(\{W_k\}\), and a set of environmental parameters \(\mathbf{E}\). The variable \(\beta\) represents the intercept, \(n\) is the path loss exponent, \(L_k\) is the loss for each wall type, \(\theta_j\) captures the contribution of each environmental factor \(E_j\), and \(k_{\text{SNR}}\) weights the effect of SNR. The term \(20 \cdot \log_{10}(f)\) accounts for frequency-dependent propagation losses, with the constant \(20\) derived from the Friis transmission equation assuming far-field conditions \cite{friisNoteSimpleTransmission1946}. The term \(\epsilon\) represents random shadowing in complex indoor environments.

\begin{figure}[hbt!]
\centering
\centerline{\includegraphics[width=.9\columnwidth]{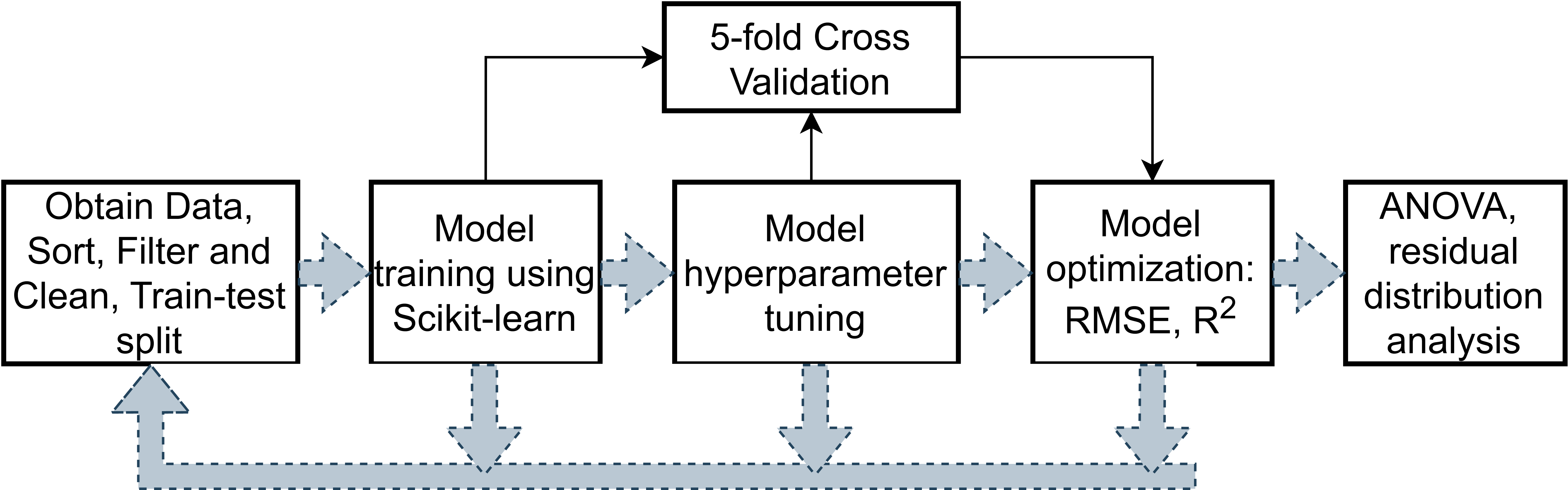}}
\caption{\label{fig:ml_pipeline} Pipeline for indoor LoRaWAN path loss modeling with iterative steps for data preparation, training, optimization, evaluation, and analysis of variance (ANOVA).}
\end{figure}

In our data analysis and modeling pipeline (Fig.~\ref{fig:ml_pipeline} ), we begin by retrieving sensor measurements from our cloud-based InfluxDB. This raw data, comprising \(1,328,334 \) field measurements of LoRaWAN metadata (RSSI, SF, SNR, etc.) and environmental parameters (temperature, humidity, barometric pressure, PM\textsubscript{2.5}, and CO\textsubscript{2} readings), is then cleaned using Python’s \texttt{Pandas} library to sort by gateway and remove duplicate records. We categorically use data transmitted by spreading factors (SFs) 7 to 10 since they balance data throughput and range, offering sufficient sensitivity for indoor propagation without excessive airtime \cite{grubelDenseIndoorSensor2022}, which can skew modeling due to extreme outliers. Furthermore, we apply isolation forests (via \texttt{Scikit-learn}'s implementation) for outlier detection and removal, ensuring data integrity before further analysis. 
\begin{figure*}[htb!]
\normalsize
\begin{equation}
PL(d, \{W_k\}, \mathbf{E}) = \beta + 10 \cdot n \cdot \log_{10}\!\Bigl(\tfrac{d}{d_0}\Bigr) + 20 \cdot \log_{10}(f) + \sum_{k=0}^{K} W_k \cdot L_k + \sum_{j=1}^{P} \theta_j \cdot E_j + k_{\text{SNR}} \cdot SNR + \epsilon,
\label{eq:pl-mw-en}
\end{equation}
\hrulefill
\end{figure*}

We use the refined dataset to represent \(\mathbf{E}\) while specifying distance \(d\) and wall parameters \(\{W_k\}\) along each signal path ( i.e., between each end device and the gateway). The dataset was split into training and testing subsets ($80\mathbin{:}20$) for fitting purposes. Initial model parameters, such as path loss exponent and wall attenuation, were selected based on domain expertise and then fine-tuned through non-linear curve fitting using the Levenberg–Marquardt algorithm. A \(5\)-fold cross-validation was applied on the entire dataset to tune hyperparameters, reduce overfitting, and ensure consistent performance across data splits. This methodology provided a robust framework for isolating the most relevant predictors and accurately approximating indoor LoRaWAN path loss in environments with multiple walls and varying environmental conditions.

\section{MLR Performance and Factor Significance}\label{sec:mlr_anova}

\subsection{MLR Model Results}\label{subsec:mlr}
\begin{table}[hbt!]
\centering
\caption{Predictor Variable Coefficients for MLR}
\label{tab:modelcoeffs}
\begin{tabular}{cccc}
\hline
\textbf{Parameter}          & \textbf{Unit}            & \textbf{ Variable} & \textbf{Coefficient} \\ 
\hline
Intercept                  & dB                       & $\beta$               & 5.435         \\ 
Path loss exponent         & -                        & $n$                     & 3.195       \\ 
Brick Wall Loss            & dB                       & $L_{\text{brick}}$      & 8.521       \\ 
Wood Wall Loss             & dB                       & $L_{\text{wood}}$       & 2.981       \\ 
CO\textsubscript{2}        & dB/ppm                   & $\theta_{\text{C}}$     & \(-0.002554 \)  \\ 
Relative humidity          & dB/\%                    & $\theta_{\text{RH}}$    & \(-0.073037\)   \\ 
PM\textsubscript{2.5}      & dB/($\mu$g/m$^3$)        & $\theta_{\text{P}}$     & \(-0.153732 \)  \\
Barometric pressure        & dB/hPa                   & $\theta_{\text{BP}}$    & \(-0.011584 \)  \\ 
Temperature                & dB/\textdegree C          & $\theta_{\text{T}}$     & \(-0.005193\)  \\ 
SNR scaling factor         & -                        & $k_{\text{SNR}}$        & \(-1.980319 \)  \\ 
\hline
\end{tabular}
\end{table}
The model coefficients in Table~\ref{tab:modelcoeffs} reflect canonical indoor propagation behavior at 868 MHz. Path loss exponents (\(n \approx 3.20\)) align with established indoor ranges (\(2\text{--}7\)) for LoRaWAN \cite{azevedoCriticalReviewPropagation2024}, indicating rapid decay from multi-wall diffraction and furniture scattering \cite{goldsmithWirelessCommunications2005}. Brick walls induce \(8.52\,\mathrm{dB}\) loss versus \(2.98\,\mathrm{dB}\) for wood, consistent with 868 MHz material attenuation studies \cite{ruddBuildingMaterialsPropagation2014}.  

Moreover, contrary to outdoor propagation models \cite{gonzalez-palacioLoRaWANPathLoss2023}, our MLR reveals negative coefficients for all environmental parameters, reflecting the distinctive influence of indoor environmental dynamics on signal propagation. This phenomenon can be attributed to office human activity. Elevated CO\textsubscript{2} and humidity often coincide with occupancy peaks, where increased human movement and HVAC operation enhance multipath richness through dynamic scattering \cite{wangXRF55RadioFrequency2024}. Temperature's modest inverse relationship may arise from thermal expansion effects that subtly alter wall material permittivity at \(868\,\mathrm{MHz}\) \cite{sebastianDielectricMaterialsWireless2010}. While such changes are minor indoors, they remain measurable and can influence boundary interactions and reflection loss. The strong SNR dependence ($k_{\mathrm{SNR}} = -1.98$) aligns with the fundamental Friis-derived receiver physics of the classical propagation theory \cite{rappaportWirelessCommunicationsPrinciples2002}, while PM\textsubscript{2.5} and pressure likely serve as spatial-temporal proxies for unmeasured human activities and HVAC state changes \cite{grubelDenseIndoorSensor2022}. Collectively, these parameters form an implicit channel state indicator that is a critical enabler for environment-aware 6G IoT networks.

Incorporating environmental parameters into the MLR framework yields a marked reduction in predictive uncertainty, with a cross-validated RMSE of $8.04\,\mathrm{dB}$ and an $R^2$ of $0.8222$ on the test set. This reflects a significant gain over the baseline model, which reported an RMSE of $10.58\,\mathrm{dB}$ and $R^2 = 0.6917$, under identical conditions. Crucially, the fraction of unexplained variance is reduced by approximately 42.3\%, indicating that environmental factors, though often overlooked in path loss modeling, play a pivotal role in capturing signal variability. This enhancement highlights improved fit and increased physical realism and generalizability, particularly in dynamic indoor environments where traditional models fall short.


\begin{figure*}[hbt!]
\centering
\includegraphics[width=.9\textwidth]{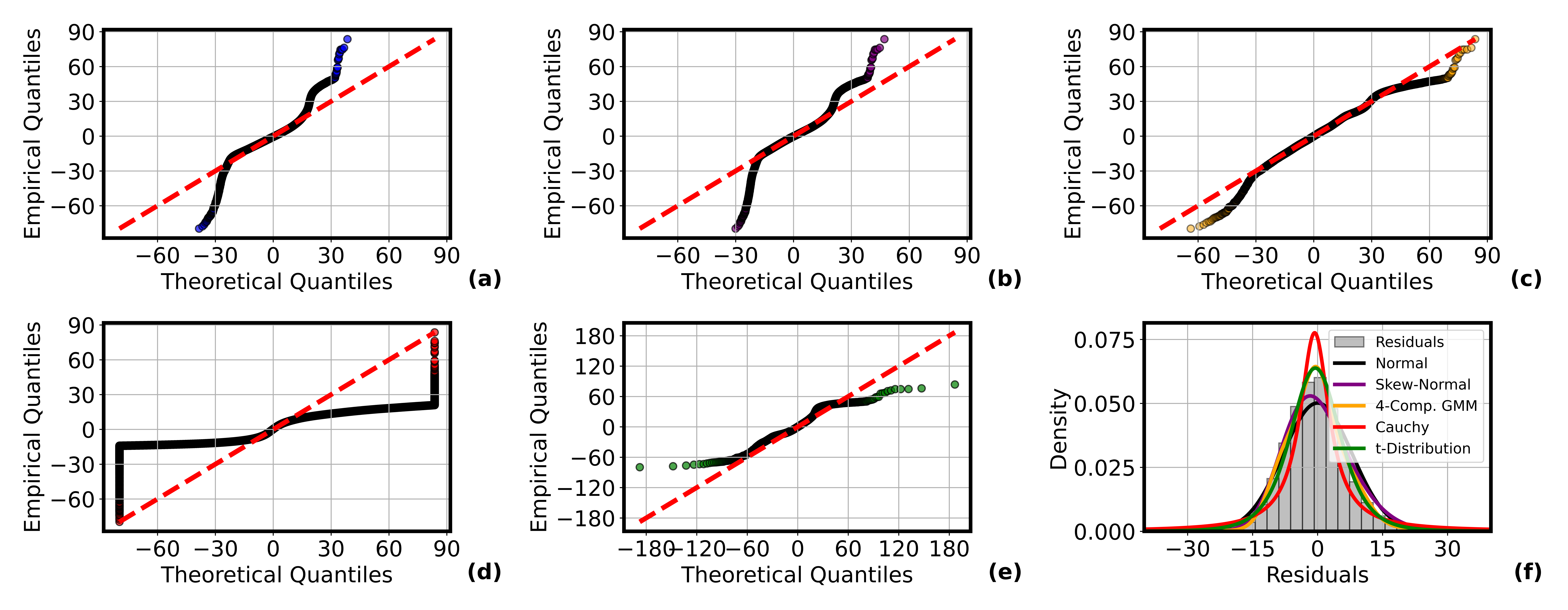}
\caption{\label{fig:qqplots} Quantile--Quantile plots comparing the OLS residual distributions of: \textbf{(a)} Normal, \textbf{(b)} Skew--Normal, \textbf{(c)} GMM (4 components), \textbf{(d)} Cauchy, and \textbf{(e)} Student's $t$. \textbf{(f)} Residual histogram overlaid with fitted probability density functions (PDFs). The four-component GMM captures distinct residual structure, modeling two central peaks (\(\mu_1, \mu_2\)) and extended tails (\(\mu_3, \mu_4\)), reflecting the multimodal and heterogeneous nature of indoor propagation environments.}

\end{figure*}


\begin{table}[hbt!]
\centering
\caption{Statistical Significance of Predictors (OLS and ANOVA)}
\label{tab:env_significance}
\begin{tabular}{ccccc}
\hline
\textbf{Parameter} & \textbf{Unit} & \textbf{\(t\)-value} & \textbf{\(F\)-statistic} & \textbf{\(p\)-value} \\
\hline
SNR & - & \(-748.46\) & \(560185.15\) & \(<0.001\) \\
Humidity & dB/\% & \(-46.41\) & \(2149.40\) & \(<0.001\) \\
CO\textsubscript{2} & dB/ppm & \(-35.54\) & \(1262.77\) & \(<0.001\) \\
PM\textsubscript{2.5} & dB/($\mu$g/m$^3$) & \(-30.06\) & \(1176.53\) & \(<0.001\) \\
Pressure & dB/hPa & \(-13.59\) & \(184.64 \)& \(<0.001\) \\
Temperature & dB/\textdegree C & \(-11.06\) & \(1.35\) & \(0.245\) \\
\hline
\end{tabular}
\end{table}

\subsection{Statistical Significance of Environmental Factors}

ANOVA tests whether each factor or factor interaction explains a significant portion of the residual variance \cite{montgomeryDesignAnalysisExperiments2017}. Predictor significance in the MLR model was evaluated through individual predictor \(t\)-tests and a Type II ANOVA to evaluate each variable's unique contribution and combined effects within the model (see Table~\ref{tab:env_significance}). All factors except temperature demonstrated highly significant effects (\(p<0.001\)), with \(t\)-values ranging from \(-13.59\) (pressure) to \(-748.46\) (SNR), and corresponding \(F\)-statistics between \(184.6\) and \(560,185\). Temperature (\(t=-11.06\), \(F=1.35\), \(p=0.245\)) had a limited marginal contribution, likely due to collinearity with other environmental parameters. Although not strictly environmental, SNR exhibited the strongest statistical significance due to its direct relationship to received signal quality. These results' strong statistical and physical coherence highlight the necessity of integrating environmentally aware variables into indoor propagation models, with their direct impacts discussed in Section ~\ref{subsec:mlr}.

\section{Shadow Fading / Residual Distribution Analysis}
\label{sec:residdist}

In addition to coefficient significance and overall model fit, we conducted a detailed residual distribution analysis to evaluate their statistical properties. Initial normality tests (Omnibus, Jarque–Bera) revealed significant deviations from Gaussian assumptions, motivating further exploration with broader distribution families that accommodate asymmetry, heavy tails, and multimodality typical of complex indoor environments. Specifically, we tested five candidate distributions: \textit{(i)} Normal, \textit{(ii)} Student's \textit{t}, and  \textit{(iii)} Cauchy \cite{johnsonContinuousUnivariateDistributions1995}, \textit{(iv)} Gaussian Mixture Model (GMM) \cite{reynoldsGaussianMixtureModels2009}, and \textit{(v)} Skew-Normal \cite{azzaliniSkewNormalRelatedFamilies2014}. GMMs are particularly suitable due to their flexibility in modeling heterogeneous subpopulations arising from dynamic indoor conditions. To determine the optimal GMM complexity without overfitting, we systematically fitted models ranging from 1 to 5 components. We then selected the best fit based on Bayesian Information Criterion (BIC), Akaike Information Criterion (AIC), and Kolmogorov–Smirnov (KS) tests.

\subsection{Goodness-of-Fit Criteria}
The ordinary least squares (OLS) regression summary reveals several key diagnostic statistics as follows: First, both the Omnibus Test (\( p < 0.001 \)) and the Jarque--Bera Test (\( p < 0.001 \)) indicate significant non-normality in the residuals. Second, the skewness of \(1.052\) and kurtosis of \(8.170\) suggest a right-heavy tail distribution with more pronounced outliers than the Normal distribution. Third, the Durbin--Watson statistic of approximately \(1.992\) suggests no strong autocorrelation is present, which reduces concerns about serial dependence in the errors. Although the model explains a large portion of the variance (\( R^2 \approx 0.825 \)), these tests emphasize that a simple Normal assumption may be insufficient to capture the error structure, especially in the tails.

We extracted the OLS residuals and fitted each distribution using Maximum Likelihood Estimation (MLE). In mathematical form, for a set of residuals \(\{r_i\}, \, i = 1, \dots, n\), and a parametric density \(f(r_i; \boldsymbol{\theta})\) parameterized by \(\boldsymbol{\theta}\), we maximize the log-likelihood in (\ref{eq:log_likelihood}):
\begin{equation}
\ell(\boldsymbol{\theta}) = \sum_{i=1}^{n} \log(f(r_i; \boldsymbol{\theta})).
\label{eq:log_likelihood}
\end{equation}
The parameter vector \(\boldsymbol{\theta}\) is estimated via the following specialized fitting routines: (i) Normal and Cauchy: (location, scale), (ii) Skew-Normal: (shape, location, scale), (iii) GMM: (means, variances, and mixture weights), and (iv) Student’s \(t\): (degrees of freedom, location, scale).
\begin{table}[hbt!]
\centering
\caption{Fit Diagnostics (Log-Likelihood, AIC, BIC, KS Test) for Residual Distributions}
\label{tab:distfits}
\begin{tabular}{@{}ccccc@{}}
\toprule
\textbf{Distribution} & \textbf{LogLik} ($\times 10^6$) & \textbf{AIC} ($\times 10^6$) & \textbf{BIC} ($\times 10^6$) & \textbf{KS stat} \\
\midrule
Normal & $-2.6330$ & $5.2660$ & $5.2660$ & $0.0570$ \\
Skew-Normal & $-2.6065$ & $5.2130$ & $5.2131$ & $0.0435$ \\
\textbf{GMM (4-comp.)} & $\mathbf{-2.5692}$ & $\mathbf{5.1384}$ & $\mathbf{5.1385}$ & $\mathbf{0.0056}$ \\
Cauchy & $-2.6716$ & $5.3432$ & $5.3432$ & $0.0779$ \\
$t$-Distribution & $-2.5808$ & $5.1626$ & $5.1626$ & $0.0210$ \\
\bottomrule
\end{tabular}
\end{table}
To objectively compare the fits, we also employed three complementary measures:  First, the log-likelihood (LL) provides a direct measure of fit, with higher values indicating a better fit under the distribution’s likelihood function. Second, the AIC \cite{akaikeNewLookStatistical1974} and BIC \cite{schwarzEstimatingDimensionModel1978} penalize model complexity to prevent overfitting. These are computed as: 
\begin{equation}
\text{AIC} = 2k - 2\ell(\hat{\boldsymbol{\theta}}),
\label{eq:aic}
\end{equation}
\begin{equation}
\text{BIC} = k \ln(n) - 2\ell(\hat{\boldsymbol{\theta}}),
\label{eq:bic}
\end{equation}
where \(k\) is the number of estimated parameters, \(n\) is the sample size, and \(\ell(\hat{\boldsymbol{\theta}})\) is the maximized log-likelihood. Both criteria penalize model complexity, but BIC applies a stronger penalty when \(n\) is large. Finally, the KS test compares each fitted cumulative distribution function (CDF) against the empirical CDF of the residuals \cite{masseyjr.KolmogorovSmirnovTestGoodness1951}. A smaller KS statistic indicates fewer discrepancies across the entire range of error values.

\subsection{Evaluating Residual Distributions}

Table~\ref{tab:distfits} summarizes the goodness-of-fit metrics, and the following observations can be made:

\subsubsection{Normal Distribution}

While computationally simple, it yields a relatively low log-likelihood (\(\approx -2.6330 \times 10^6\)) and higher AIC (\(\approx 5.2660 \times 10^6\)). The KS statistic (\(\approx 0.0570\)) and visual Q--Q plots (Fig. \ref{fig:qqplots}(a)) confirm that the Normal distribution underestimates the heavier tails.

\subsubsection{Skew-Normal Distribution}

By introducing a skewness parameter, this model improved the tail fit slightly over the pure Gaussian but still underperformed for large positive residuals (Fig. \ref{fig:qqplots}(b)). Its AIC/BIC was better than the Normal (\(\approx 5.2130 \times 10^6\)) but insufficient to unseat the GMM.

\subsubsection{Gaussian Mixture Model (GMM)}

A four-component GMM achieved the highest log-likelihood (\(-2.5692\times10^6\)), lowest AIC (\(5.1384\times10^6\)), and smallest KS statistic (\(0.0056\)) among all tested distributions, clearly outperforming simpler single-distribution approaches. Visual inspection (Fig.~\ref{fig:qqplots}(c)) further confirmed that the four-component mixture closely matched empirical residual distributions, effectively capturing multiple overlapping error-generating processes typical of dynamic indoor environments influenced by occupancy variations(low/high), structural complexity, and environmental fluctuations (e.g., HVAC cycles).

\subsubsection{Cauchy Distribution}

Although it handles extreme outliers via its heavy tails (Fig. \ref{fig:qqplots}(d)), it misrepresents the bulk of the distribution, as evidenced by its relatively high KS value (\(\approx 0.0779\)). Its higher AIC/BIC (\(\approx 5.3432 \times 10^6\)) further supports that it is not well-suited for these residuals.

\subsubsection[Student's t-Distribution]{Student's \(t\)-Distribution}

This distribution offers thicker tails than the Normal and achieves a moderate fit (KS \(\approx 0.0210\); AIC \(\approx 5.1626 \times 10^6\)). However, the upper tail in Fig. \ref{fig:qqplots}(e) still shows a noticeable deviation from the theoretical line, indicating that even flexible unimodal distributions can fall short when the data arise from multiple latent processes.

\subsubsection{Residual Distribution}

Fig. \ref{fig:qqplots}(f) further illustrates how the GMM’s density curve aligns closely with the empirical histogram, particularly around the primary peak near zero and in the heavier tails. The strong performance of the GMM suggests that residuals from indoor LoRaWAN power loss models may arise from at least four subpopulations, possibly linked to different building layouts, occupant densities, or microclimatic variations. This multimodal behavior is not captured by the other standard unimodal families compared. 


\section{Conclusion and Future Directions}
\label{sec:conclusion}

This work introduced an integrated approach for indoor LoRaWAN path loss modeling that combines multiple linear regression (MLR), ANOVA-based significance testing, and multimodal residual (shadow fading) analysis. Utilizing a dataset of \(1,328,334 \) field measurements, we demonstrated that incorporating environmental factors (relative humidity, temperature, barometric pressure, particulate matter, and carbon dioxide) reduced unexplained variance by $42.3\%$. A key finding is that a four-component Gaussian Mixture Model (GMM) accurately reflects heterogeneous error patterns, likely driven by fluctuating occupancy or microclimatic conditions. Therefore, the proposed environment-aware strategy holds promise for more resilient indoor LoRaWAN deployments, aligning with the needs of 6G and IoT systems that demand consistently robust network performance and energy efficiency.

While our primary analysis relied on the simplicity and interpretability of MLR, inspection of the residuals revealed signs of nonlinearity and complex interactions among predictors that warrant further exploration. In future work, we plan to extend our analysis by incorporating advanced nonlinear machine learning techniques (e.g., ensemble methods and deep neural architectures) to capture latent patterns in indoor propagation environments comprehensively. These methods could offer improved prediction accuracy while maintaining robust statistical validations when coupled with real-time occupancy monitoring. 

\section*{Data Availability}
The dataset and analysis scripts used for this work are openly available at: \url{https://github.com/nahshonmokua/LoRaWAN-IndoorPathLossModeling-ML}

\section*{Acknowledgment}
The authors thank the Ubiquitous Computing Group at the University of Siegen for their support and contribution during the data collection campaign.

\bibliographystyle{IEEEtran}
\bibliography{refs}

\begin{thebibliography}{10}
\providecommand{\url}[1]{#1}
\csname url@samestyle\endcsname
\providecommand{\newblock}{\relax}
\providecommand{\bibinfo}[2]{#2}
\providecommand{\BIBentrySTDinterwordspacing}{\spaceskip=0pt\relax}
\providecommand{\BIBentryALTinterwordstretchfactor}{4}
\providecommand{\BIBentryALTinterwordspacing}{\spaceskip=\fontdimen2\font plus
\BIBentryALTinterwordstretchfactor\fontdimen3\font minus \fontdimen4\font\relax}
\providecommand{\BIBforeignlanguage}[2]{{%
\expandafter\ifx\csname l@#1\endcsname\relax
\typeout{** WARNING: IEEEtran.bst: No hyphenation pattern has been}%
\typeout{** loaded for the language `#1'. Using the pattern for}%
\typeout{** the default language instead.}%
\else
\language=\csname l@#1\endcsname
\fi
#2}}
\providecommand{\BIBdecl}{\relax}
\BIBdecl

\bibitem{cattaniExperimentalEvaluationReliability2017}
M.~Cattani, C.~A. Boano, and K.~R{\"o}mer, ``An {{Experimental Evaluation}} of the {{Reliability}} of {{LoRa Long-Range Low-Power Wireless Communication}},'' \emph{Journal of Sensor and Actuator Networks}, vol.~6, no.~2, p.~7, Jun. 2017.

\bibitem{grubelDenseIndoorSensor2022}
J.~Gr{\"u}bel, T.~Thrash, L.~Aguilar, M.~{Gath-Morad}, D.~H{\'e}lal, R.~W. Sumner, C.~H{\"o}lscher, and V.~R. Schinazi, ``Dense {{Indoor Sensor Networks}}: {{Towards}} passively sensing human presence with {{LoRaWAN}},'' \emph{Pervasive and Mobile Computing}, vol.~84, p. 101640, Aug. 2022.

\bibitem{xuMeasurementCharacterizationModeling2020}
W.~Xu, J.~Y. Kim, W.~Huang, S.~S. Kanhere, S.~K. Jha, and W.~Hu, ``Measurement, {{Characterization}}, and {{Modeling}} of {{LoRa Technology}} in {{Multifloor Buildings}},'' \emph{IEEE Internet of Things Journal}, vol.~7, no.~1, pp. 298--310, Jan. 2020.

\bibitem{gonzalez-palacioLoRaWANPathLoss2023}
M.~{Gonz{\'a}lez-Palacio}, D.~{Tob{\'o}n-Vallejo}, L.~M. {Sep{\'u}lveda-Cano}, S.~R{\'u}a, G.~Pau, and L.~B. Le, ``{{LoRaWAN Path Loss Measurements}} in an {{Urban Scenario}} including {{Environmental Effects}},'' \emph{Data}, vol.~8, no.~1, p.~4, Jan. 2023.

\bibitem{kimExperiencingLoRaNetwork2019}
D.-H. Kim, E.-K. Lee, and J.~Kim, ``Experiencing {{LoRa Network Establishment}} on a {{Smart Energy Campus Testbed}},'' \emph{Sustainability}, vol.~11, no.~7, p. 1917, Jan. 2019.

\bibitem{mokuaobiriComprehensiveDataDescription2025}
N.~Mokua~Obiri and K.~{van Laerhoven}, ``A {{Comprehensive Data Description}} for {{LoRaWAN Path Loss Measurements}} in an {{Indoor Office Setting}}: {{Effects}} of {{Environmental Factors}},'' \emph{IEEE Access}, vol.~13, pp. 83\,148--83\,170, 2025.

\bibitem{gonzalez-palacioMachineLearningBasedCombinedPath2023a}
M.~{Gonz{\'a}lez-Palacio}, D.~{Tob{\'o}n-Vallejo}, L.~M. {Sep{\'u}lveda-Cano}, S.~R{\'u}a, and L.~B. Le, ``Machine-{{Learning-Based Combined Path Loss}} and {{Shadowing Model}} in {{LoRaWAN}} for {{Energy Efficiency Enhancement}},'' \emph{IEEE Internet of Things Journal}, vol.~10, no.~12, pp. 10\,725--10\,739, Jun. 2023.

\bibitem{azevedoCriticalReviewPropagation2024}
J.~A. Azevedo and F.~Mendon{\c c}a, ``A {{Critical Review}} of the {{Propagation Models Employed}} in {{LoRa Systems}},'' \emph{Sensors}, vol.~24, no.~12, p. 3877, Jan. 2024.

\bibitem{robles-encisoLoRaZigbee5G2023}
R.~{Robles-Enciso}, I.~P. {Morales-Arag{\'o}n}, A.~{Serna-Sabater}, M.~T. {Mart{\'i}nez-Ingl{\'e}s}, A.~{Mateo-Aroca}, J.-M. {Molina-Garcia-Pardo}, and L.~{Juan-Ll{\'a}cer}, ``{{LoRa}}, {{Zigbee}} and {{5G Propagation}} and {{Transmission Performance}} in an {{Indoor Environment}} at 868 {{MHz}},'' \emph{Sensors}, vol.~23, no.~6, p. 3283, Jan. 2023.

\bibitem{aksoyComparativeAnalysisEnd2024}
A.~Aksoy, {\"O}.~Y{\i}ld{\i}z, and S.~E. Karl{\i}k, ``Comparative {{Analysis}} of {{End Device}} and {{Field Test Device Measurements}} for {{RSSI}}, {{SNR}} and {{SF Performance Parameters}} in an {{Indoor LoRaWAN Network}},'' \emph{Wireless Personal Communications}, vol. 134, no.~1, pp. 339--360, Jan. 2024.

\bibitem{alkhazmiAnalysisRealWorldLoRaWAN2023}
E.~H. Alkhazmi, S.~M. Elkawafi, A.~A. Aldarrat, M.~A. Abbas, H.~Abubakr, and H.~A. Shamatah, ``Analysis of {{Real-World LoRaWAN Network Performance Across Outdoor}} and {{Indoor Scenarios}},'' in \emph{2023 {{IEEE}} 11th {{International Conference}} on {{Systems}} and {{Control}} ({{ICSC}})}, Dec. 2023, pp. 329--334.

\bibitem{muppalaInvestigationIndoorLoRaWAN2021}
R.~Muppala, A.~Navnit, S.~Poondla, and A.~M. Hussain, ``Investigation of {{Indoor LoRaWAN Signal Propagation}} for {{Real-World Applications}},'' in \emph{2021 6th {{International Conference}} for {{Convergence}} in {{Technology}} ({{I2CT}})}, Apr. 2021, pp. 1--5.

\bibitem{harindaPerformanceLiveMultiGateway2022}
E.~Harinda, A.~J. Wixted, A.-U.-H. Qureshi, H.~Larijani, and R.~M. Gibson, ``Performance of a {{Live Multi-Gateway LoRaWAN}} and {{Interference Measurement}} across {{Indoor}} and {{Outdoor Localities}},'' \emph{Computers}, vol.~11, no.~2, p.~25, Feb. 2022.

\bibitem{voAdvancePathLoss2024}
H.~Vo, V.~Hoang Long~Nguyen, V.~L. Tran, F.~Ferrero, F.-Y. Lee, and M.-H. Tsai, ``Advance {{Path Loss Model}} for {{Distance Estimation Using LoRaWAN Network}}'s {{Received Signal Strength Indicator}} ({{RSSI}}),'' \emph{IEEE Access}, vol.~12, pp. 83\,205--83\,216, 2024.

\bibitem{montgomeryIntroductionLinearRegression2012}
D.~C. Montgomery, E.~A. Peck, and G.~G. Vining, \emph{Introduction to {{Linear Regression Analysis}}}.\hskip 1em plus 0.5em minus 0.4em\relax John Wiley \& Sons, Apr. 2012.

\bibitem{farawayPracticalRegressionAnova2002}
J.~J. Faraway, \emph{Practical {{Regression}} and {{Anova}} Using {{R}}}.\hskip 1em plus 0.5em minus 0.4em\relax University of Bath Bath, 2002.

\bibitem{europeancommissionCOSTAction2311999}
{European Commission}, ``{{COST Action}} 231 - {{Digital}} mobile radio towards future generation systems,'' {European Commission, Directorate-General for the Information Society and Media}, Luxembourg, Final {{Report}}, 1999.

\bibitem{friisNoteSimpleTransmission1946}
H.~Friis, ``A {{Note}} on a {{Simple Transmission Formula}},'' \emph{Proceedings of the IRE}, vol.~34, no.~5, pp. 254--256, May 1946.

\bibitem{goldsmithWirelessCommunications2005}
A.~Goldsmith, \emph{Wireless {{Communications}}}.\hskip 1em plus 0.5em minus 0.4em\relax Cambridge University Press, Aug. 2005.

\bibitem{ruddBuildingMaterialsPropagation2014}
R.~Rudd, K.~Craig, M.~Ganley, and R.~Hartless, ``Building {{Materials}} and {{Propagation}},'' Ofcom, United Kingdom, Final {{Report}} 2604/BMEM/R/3/2.0, 2014.

\bibitem{wangXRF55RadioFrequency2024}
F.~Wang, Y.~Lv, M.~Zhu, H.~Ding, and J.~Han, ``{{XRF55}}: {{A Radio Frequency Dataset}} for {{Human Indoor Action Analysis}},'' \emph{Proc. ACM Interact. Mob. Wearable Ubiquitous Technol.}, vol.~8, no.~1, pp. 21:1--21:34, Mar. 2024.

\bibitem{sebastianDielectricMaterialsWireless2010}
M.~T. Sebastian, \emph{Dielectric {{Materials}} for {{Wireless Communication}}}.\hskip 1em plus 0.5em minus 0.4em\relax Elsevier, Jul. 2010.

\bibitem{rappaportWirelessCommunicationsPrinciples2002}
T.~S. Rappaport, \emph{Wireless Communications: {{Principles}} and Practice}, ser. Prentice {{Hall}} Communications Engineering and Emerging Technologies Series.\hskip 1em plus 0.5em minus 0.4em\relax Upper Saddle River, N.J: Prentice Hall, 2002, includes bibliographical references and index.

\bibitem{montgomeryDesignAnalysisExperiments2017}
D.~C. Montgomery, \emph{Design and {{Analysis}} of {{Experiments}}}.\hskip 1em plus 0.5em minus 0.4em\relax John Wiley \& Sons, 2017.

\bibitem{johnsonContinuousUnivariateDistributions1995}
N.~L. Johnson, S.~Kotz, and N.~Balakrishnan, \emph{Continuous {{Univariate Distributions}}, {{Volume}} 2}.\hskip 1em plus 0.5em minus 0.4em\relax John Wiley \& Sons, May 1995.

\bibitem{reynoldsGaussianMixtureModels2009}
D.~Reynolds, ``Gaussian {{Mixture Models}},'' in \emph{Encyclopedia of {{Biometrics}}}, S.~Z. Li and A.~Jain, Eds.\hskip 1em plus 0.5em minus 0.4em\relax Boston, MA: Springer US, 2009, pp. 659--663.

\bibitem{azzaliniSkewNormalRelatedFamilies2014}
A.~Azzalini and A.~Capitanio, \emph{The {{Skew-Normal}} and {{Related Families}}}.\hskip 1em plus 0.5em minus 0.4em\relax Cambridge University Press, 2014.

\bibitem{akaikeNewLookStatistical1974}
H.~Akaike, ``A new look at the statistical model identification,'' \emph{IEEE Transactions on Automatic Control}, vol.~19, no.~6, pp. 716--723, Dec. 1974.

\bibitem{schwarzEstimatingDimensionModel1978}
G.~Schwarz, ``Estimating the {{Dimension}} of a {{Model}},'' \emph{The Annals of Statistics}, vol.~6, no.~2, pp. 461--464, Mar. 1978.

\bibitem{masseyjr.KolmogorovSmirnovTestGoodness1951}
F.~J. Massey~Jr., ``The {{Kolmogorov-Smirnov Test}} for {{Goodness}} of {{Fit}},'' \emph{Journal of the American Statistical Association}, vol.~46, no. 253, pp. 68--78, Mar. 1951.

\end{thebibliography}

\end{document}